# Theoretical Modeling of Internal Hydraulic Jump in Density Currents

Bahar Firoozabadi[1], Milad Samie[1], Asghar Aryanfar[2] and Hossein Afshin[1]

[1]*Mechanical Engineering Department, Sharif University of Technology, Tehran, Iran.*
[2]*Mechanical and Civil Engineering Department, California Institute of Technology, Pasadena, CA.*

**ABSTRACT**

In this paper, we propose an analytical framework for internal hydraulic jumps. Density jumps or internal hydraulic jumps occur when a supper critical flow of water discharges into a stagnant layer of water with slightly different density. The approach used here is control volume method which is also used to analyze ordinary hydraulic jumps. The important difference here is that entrainment is taken into account. Using conservation equations with the aid of some simplifying assumptions we come to an equation that gives jump downstream height as function of jump upstream characteristics and the entrainment. To determine the magnitude of downstream height we use an experimental equation for calculating the entrainment. Finally we verify our framework by comparing the height that we gain from the derived equation with some experimental data.

## INTRODUCTION

Performing an experiment in the lab on density flows, a two-layer flow of fluids with slightly different densities, a hydraulic jump on slope entrance of the experiment setup was observed in some specific conditions. This jump is called a "density jump" in which when the fluids are miscible there is usually entrainment from ambient flow into the dense layer due to density difference, changing the density of the flow across the jump. Density jumps occur when a heavy fluid with super critical velocity discharges under a lighter fluid and is seen in the ocean, on mountains when an avalanche happens and in hydraulic structures.

Many efforts have been made to create an analytical framework for density jumps. Yih & Guha (1955) assumed that there is no entrainment in the internal jump and neglecting the friction forces they could solve the problem completely.

By taking the entrainment into account and neglecting the friction forces Macango & Macango (1975) needed an additional relation. To reach it they assumed that the energy loss in the internal jump is a fraction of energy depletion in internal jumps with no entrainment (Yih & Guha theory), i.e. $\alpha \Delta E$ where $\Delta E$ is the energy loss in jump with no entrainment and $\alpha$ is a factor gained experimentally. They also assumed that the entrainment occurred at the foot of the jump.

Chu & Baddour (1977) used a different suggestion. In a hydraulic jump in two-layered flow, expansion occurs in the lower layer while the upper one contracts. By putting dye in the upper layer near the contact surface they observed no mixing while the dye put in the contracting layer diffused considerably. Using this observation they assumed that the energy loss in the contracting layer is insignificant with respect to that of the expanding layer so they utilized energy conservation for upper layer and momentum equation for the system of two flows which enabled them to solve the problem comprehensively.

Wood & Simpson (1983) used Chu & Baddour method. Using this method, they analyzed several cases: the jump advancing into two stationary layers, the jump behind the moving obstacle and the stationary entraining internal jumps in a duct. In the last case they reached a set of equations which were not solved to yield an explicit equation. They solved the equations for the extreme case of very deep duct relating $Fr_2$ to $\dfrac{Q_2}{Q_1}$ and plotted it.

Wilkinson and Wood studied density jumps theoretically and experimentally, with focus on the density jump being controlled by a crest wire. They suggested that the jump is divided into two regions: The entrainment region and the roller region in which there is no entrainment. They showed that to solve the problem entirely an additional boundary condition is needed and with the assumption of a critical flow over the weir downstream of the jump and using the conversation of energy equation they provided the additional boundary condition.

None of the above studies yield to an explicit equation for density jumps on slopes which is the main work done in this study.

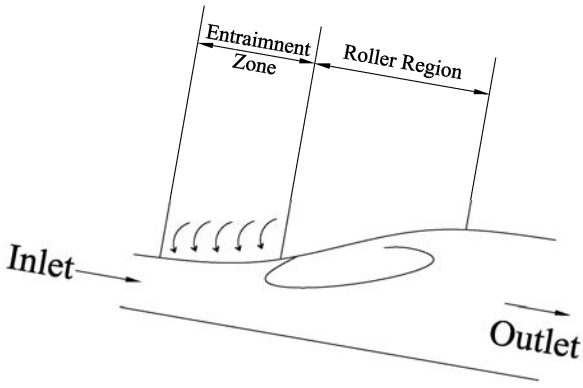

Figure 1. Internal jump in density currents showing its regions

**THEORY**

The approach used in this study is control volume analysis. As a density jump like hydraulic jump is absolutely turbulent, energy is not conserved in it, though we can use momentum relation for the control volume.

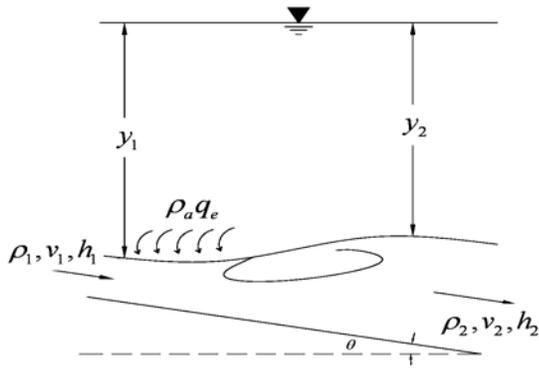

Figure 2. Internal jump in density currents showing involved parameters.

Mass conservation equation gives:

$$\rho_1 v_1 h_1 + \rho_a q_e = \rho_2 v_2 h_2 \tag{1}$$

This can be written in the form:

$$1 + \varepsilon = \frac{\rho_2 v_2 h_2}{\rho_1 v_1 h_1} = \frac{Q_2}{Q_1} \tag{2}$$

$\varepsilon$ in the Eq.1 represents entrainment and is defined as the ratio of mass flux of ambient water to that of upstream water:

$$\varepsilon = \frac{\rho_a q_e}{\rho_1 v_1 h_1}$$

Fluid density is related to salinity by the following equation:

$$\rho = \rho_o (1 + \beta s) \tag{3}$$

Eq.3 resembles to more familiar equation relating density to temperature:

$$\rho = \rho_o (1 + \beta \Delta \theta)$$

Using Eq.1 with the aid of Eq.3 yields:

$$(v_1 h_1 + q_e - v_2 h_2) + \beta(s_1 v_1 h_1 + s_a q_e) = \beta s_2 v_2 h_2 \tag{4}$$

Assuming that the upstream water is sufficiently dilute the first parenthesis in Eq.4 is equal to zero so the equation reduces to:

$$s_1 v_1 h_1 + s_a q_e = s_2 v_2 h_2 \tag{5}$$

Starting with Eq.1' and using Eq.5 it can be seen that:

$$\Delta \rho_2 = \frac{\Delta \rho_1 v_1 h_1}{v_1 h_1 + q_e} \tag{6}$$

Where:

$$\Delta \rho_i = \rho_i - \rho_a$$

To derive momentum conservation equation one must first figure out forces which are pressure and gravity forces shown in fig.3. Assuming that both the shear stress between the layers and interfacial slope are negligible the forces are derived.

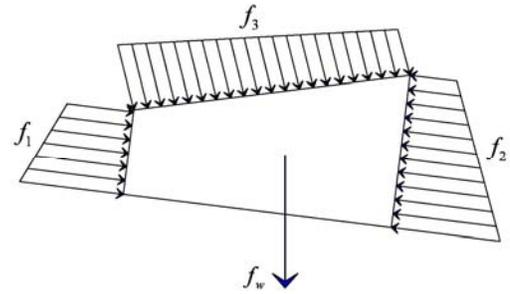

Figure 3: Free body diagram of control volume

The forces due to pressure as shown on the fig.3 are:

$$f_1 = \frac{\rho_1 g h_1^2}{2} + \rho_a g y_1 h_1 \tag{6-a}$$

$$f_2 = \frac{\rho_2 g h_2^2}{2} + \rho_a g y_2 h_2 \tag{6-b}$$

$$f_3 = \rho_a g \frac{y_1 + y_2}{2}(h_1 - h_2)$$
$$= \frac{\rho_a g}{2}(y_1 h_2 - y_1 h_1 + y_2 h_2 - y_2 h_1) \quad (6\text{-c})$$

Using a geometric analysis one obtains:

$$y_1 - y_2 - l\sin\theta = (h_2 - h_1)\cos\theta \quad (7)$$

Substituting Eq.7 into Eq.6-c, we have:

$$f_3 = \frac{\rho_a g}{2}(h_1 + h_2)((h_2 - h_1)\cos\theta - l\sin\theta) \quad (6\text{-d})$$

The jump occurs on a slope so we have a force due to gravity. Assuming that the term (ρ1-ρa) is sufficiently small and taking a trapezoidal shape for the control volume, gravity force will be:

$$f_w = \rho_a g \frac{h_1 + h_2}{2} l \sin\theta \quad (8)$$

Using momentum conservation equation with the aid of Eqs.6 and Eq.8, it is found that:

$$\frac{gh_1^2}{2}\Delta\rho_1 - \frac{gh_2^2}{2}\Delta\rho_2 + \frac{\rho_a g}{2}(h_2^2 - h_1^2)(\cos\theta - 1)$$
$$= \frac{Q_2^2}{\rho_a h_1} - \frac{Q_1^2}{\rho_a h_2} \quad (9)$$

In writing the Eq.9 assuming that the upstream water is sufficiently dilute, $\rho_2$ and $\rho_1$ have been taken equal to $\rho_a$ on the right side of the equation.

Introducing Eq.1 and Eq.5 into the momentum Eq. 9, yields:

$$(\frac{h_2}{h_1})^3 (1 - \frac{\rho_a}{\Delta\rho_1}(1+\varepsilon)(\cos\theta - 1))$$
$$-(\frac{h_2}{h_1})(1+\varepsilon)(2Fr_1^2 + 1 + \frac{\rho_a}{\Delta\rho_1}(\cos\theta - 1)) \quad (10)$$
$$+ 2Fr_1^2(1+\varepsilon)^3 = 0$$

Where $Fr_1$, the upstream Froude number is defined as:

$$Fr_1 = \frac{v_1}{\sqrt{(\Delta\rho_1/\rho_1)gh_1}}$$

For $\theta = 0$, $\varepsilon = 0$ and $\Delta\rho = \rho$, Eq.10 reduces to the familiar equation of ordinary hydraulic jump:

$$(\frac{h_2}{h_1})^3 - (\frac{h_2}{h_1})(2Fr_1^2 + 1) + 2Fr_1^2 = 0 \quad (11)$$

This equation has one physically valid solution:

$$\frac{h_2}{h_1} = \frac{\sqrt{1 + 8Fr_1^2} - 1}{2} \quad (12)$$

Eq.10 is more complex since θ and ε are involved in it, and has the following solution:

$$\frac{h_2}{h_1} = 2\sqrt{n}\cos(u/3) \quad (13)$$

Where:

$$\cos(u) = \frac{m}{n\sqrt{n}}$$

$$n = \frac{1}{3} \cdot \frac{(1+\varepsilon)(2Fr_1^2 + 1 + \frac{\rho_a}{\Delta\rho_1}(\cos\theta - 1))}{(1 - \frac{\rho_a}{\Delta\rho_1}(1+\varepsilon)(\cos\theta - 1))},$$

$$m = -\frac{Fr_1^2(1+\varepsilon)^3}{(1 - \frac{\rho_a}{\Delta\rho_1}(1+\varepsilon)(\cos\theta - 1))} \quad (14)$$

The entrainment, ε, has a maximum value according to Eq. 13 as $\frac{h_2}{h_1}$ has, and it can be seen in the Fig.3 which shows the ratio $\frac{h_2}{h_1}$ as function of $\varepsilon$, that for various Froude numbers the ratio $\frac{h_2}{h_1}$ rises as $\varepsilon$ increases and then falls. Due to Fig.3 there are two solutions for one entrainment. The smaller one stands for $h_2'$ and the larger represents $h_2$.

Note that in density jumps there are three unknown variables: $h_2$, $v_2$ and $\rho_2$. To find them one needs three independent equations. As energy is not conserved we have mass and momentum relation which are not enough for solving the problem explicitly. We used an assumption and with it we provided an additional equation. Supposing that the density difference is sufficiently small for the ambient, upstream and downstream density to be equal so as to change mass flux relation into volume flux, we reached Eq.5 that was the key to our solution.

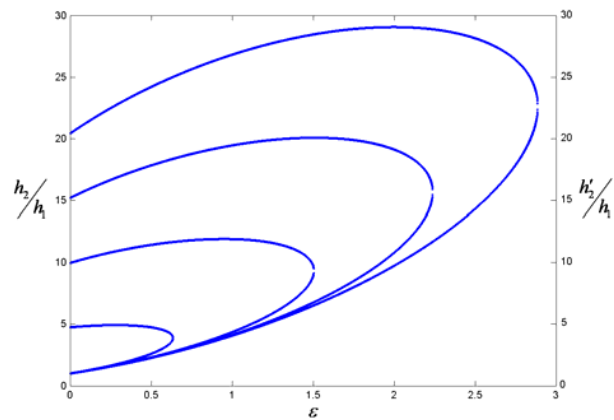

Figure 4. The ratio of height after entrainment zone $h_2'$ and height after jump $h_2$ to initial height $h_1$ versus entrainment $\varepsilon$.

**EQUATION VERIFYING**

In order to verify Eq.10 we used data that Garcia (1993) obtained in an experiment on density currents. First, we needed to measure the entrainment in that Eq.10 Gives the value of $\frac{h_2}{h_1}$ as function of entrainment. Parker et al. proposed an equation for entrainment for turbidity currents which can be used for density currents with reasonable approximation:

$$\varepsilon = \frac{0.00153}{0.0204 + Ri}$$
(15)

Where $Ri$ denotes bulk Richardson number which is related to Froude number as

$$Ri = \frac{1}{Fr^2}$$
(16)

Using the upstream conditions for a typical run by Garcia with the aid of parker equation, Eq.10 gave $\frac{h_2}{h_1}$ for each condition which is sorted in table 1 comparing with Garcia observations. The obtained data is illustrated in Fig.3.

| No. | $\rho_1$ $(kg/m^3)$ | $c_1$ $(kg/m^3)$ | $h_2/h_1$ (G-93) | $h_2/h_1$ (Eq.10) | Err (%) |
|---|---|---|---|---|---|
| SAL26 | 1002.1 | 0.21 | 2 | 2.62 | 24 |
| SAL27 | 1004 | 0.23 | 2.05 | 2.49 | 18 |
| SAL28 | 1008 | 0.25 | 2.14 | 2.37 | 10 |
| SAL29 | 1012 | 0.27 | 2.2 | 2.26 | 3 |
| SAL11 | 1013 | 0.25 | 2.36 | 2.37 | 0 |

Table 1. Downstream height from Garcia vision and theory for density currents for 5 upstream conditions. ($\theta = 0$)

The error is calculated through the following equation:

$$Error = \frac{Eq.10\ Val. - Garcia\ Val.}{Eq.10\ Val.} \times 100$$

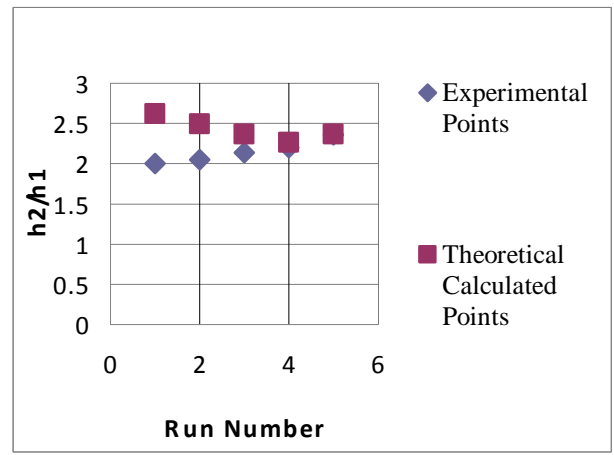

Figure 5. Comparison between experimental downstream heights *from Garcia (1993) and theory for density currents.*

Eq.10 also performs well for turbidity currents. Using Eq.10 for experimental data from Garcia (1993) for turbidity currents utilizing Eq.15 for entrainment determination the following results were obtained. The results are sorted in table 2 and illustrated in Fig.4. As it can be seen the theory results are acceptably close to experimental data.

The error calculation is the same as Table 1.

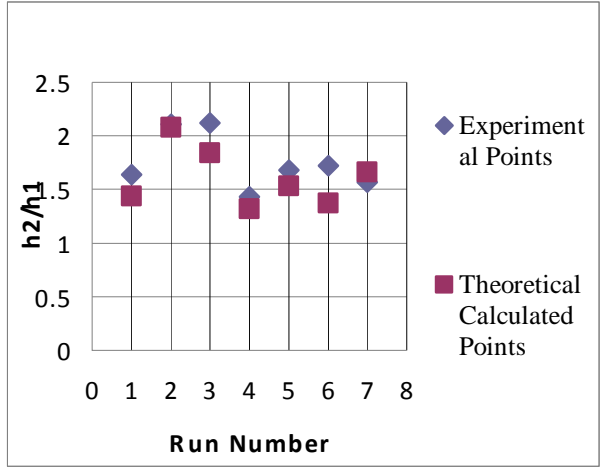

## CONCLUSION

In this paper we modeled density jumps occurred in density currents using control volume approach on slopes and regarding density difference. We obtained an equation that by solving it we can theoretically find the downstream jump height ($h_2$) as function of upstream Froude number, slope angle, and water entrainment. We also confirmed our model by comparing the results gained by the experimental results from Garcia model (1993). As there was close correlation between the analytical and experimental results and their differences were insignificant, our model proved to perform well for both density currents and turbidity currents, although it had been derived for density currents.

*Table 2. Downstream height from Garcia vision and theory for turbidity currents for 7 upstream conditions. ($\theta = 0$)*

## NOMENCLATURE

$f_1$ upstream pressure force of the jump;

$f_2$ downstream pressure force of the jump;

$f_3$ upper surface pressure force;

$f_w$ body force;

$Fr$ Froude number;

$Fr_1$ upstream Froude number of the jump;

$Fr_2$ downstream Froude number of the jump;

$g$ gravity accelerations

$h_1$ upstream height of the jump;

$h_2$ downstream height of the jump;

$h_2'$ entrainment zone downstream height;

$l$ jump length;

$m$ a mathematical coefficient introduced in the text;

$n$ a mathematical coefficient introduced in the text;

$q_e$ entrainment volume flux;

$Q_1$ upstream mass flux of the jump;

$Q_2$ downstream mass flux of the jump;

$Ri$ bulk Richardson number;

$s$ water salinity;

$s_1$ upstream water salinity of the jump;

$s_2$ downstream water salinity of the jump;

$s_a$ ambient water salinity;

$u$ a mathematical coefficient introduced in the text;

| Number | $\rho_1 (kg/m^3)$ | $c_i (kg/m^3)$ | $h_2/h_1$ (G-93) | $h_2/h_1$ (Eq.10) | Err (%) |
|---|---|---|---|---|---|
| NOVA1 | 1001.3 | 0.56 | 1.64 | 1.44 | 12 |
| NOVA2 | 1002.48 | 0.31 | 2.11 | 2.08 | 1 |
| NOVA3 | 1002.48 | 0.38 | 2.12 | 1.84 | 13 |
| DAPER1 | 1001.43 | 0.64 | 1.43 | 1.32 | 8 |
| DAPER2 | 1001.33 | 0.51 | 1.68 | 1.53 | 9 |
| DAPER4 | 1002.95 | 0.61 | 1.72 | 1.37 | 20 |
| DAPER5 | 1004.29 | 0.45 | 1.57 | 1.66 | 6 |

$v_1$ upstream water mean velocity of the jump;

$v_2$ downstream water mean velocity of the jump;

$y_1$ distance between the upper edge of jump upstream height and ambient water surface.

$y_2$ distance between the upper edge of jump downstream height and ambient water surface.

$\rho_0$ initial water density;

$\rho_1$ upstream water density of the jump;

$\rho_2$ downstream water density of the jump;

$\rho_a$ ambient water density;

$\varepsilon$ entrainment coefficient;

$\theta$ slope angle;

$\beta$ water salinity coefficient;